\documentclass[preprint,showpacs,preprintnumbers,amsmath,amssymb, superscriptaddress,longbibliography,nofootinbib]{revtex4-1}
\usepackage{graphicx} 
\usepackage{xcolor}
\usepackage{amsmath}
 \usepackage{hyperref}
\usepackage[top = 2cm, bottom = 2cm, right = 2cm, left =2cm]{geometry}

\begin{document}

\title{Pure Lovelock Gravity regular black holes}

\author{Milko Estrada }
\email{milko.estrada@gmail.com}
\affiliation{Facultad de Ingeniería y Empresa, Universidad Católica Silva Henríquez, Chile}

\author{Rodrigo Aros}
\email{raros@unab.cl}
\affiliation{Departamento de Ciencias Fisicas, Universidad Andres Bello, Av. Republica 252, Santiago,Chile}

\date{\today}

\begin{abstract}
We present a new family of regular black holes (RBH) in Pure Lovelock gravity, where the energy density is determined by the gravitational vacuum tension, which varies for each value of $n$ in each Lovelock case. Speculatively, our model may capture quantum effects through gravitational tension. In this way, a hypothetical analogy is drawn between the pair production ratio in the Schwinger effect and our energy density. A notable feature of our model is that the regular solution closely resembles the vacuum solution before reaching the event horizon. For odd $n$, the transverse geometry is spherical, with phase transitions occurring during evaporation, and the final state of this process is a remnant. For even $n$, the transverse geometry is non trivial and corresponds to a hyperboloid. In the case of $d=2n+1$ with even $n$, we find an RBH without a dS core and no inner horizon (whose presence has been recently debated in the literature due to the question of whether its presence is unstable or not), and no phase transitions. For $d > 2n + 1$ with even $n$, the RBH possesses both an event horizon and a cosmological horizon, also with no inner horizon present. The existence of the cosmological horizon arises without the usual requirement of a positive cosmological constant. From both numerical and analytical analysis, we deduce that as the event horizon expands and the cosmological horizon contracts, thermodynamic equilibrium is achieved in a remnant when the two horizons coincide.
\end{abstract}

\maketitle

\section{Introduction}

In recent years, several events, such as the discovery of gravitational waves from the collision of two rotating black holes \cite{LIGOScientific:2016aoc}, have positioned General Relativity as an effective theory for describing gravitational phenomena. 

Nowadays, it is well-accepted that black hole physics and gravity are far from being finished and several core aspects are still unclear. For instance, since the original works by Penrose and Hawking, it is well-accepted that, considering only the known quantum phenomena, gravitational collapse could lead to black hole solutions containing singularities. In this connection, it is well established that the infinite tidal forces near a black hole’s singularity can lead to the infinite stretching of an object, a phenomenon known as spaghettification. See for example the introduction of \cite{Hong:2020bdb}. However, this prediction now is sometimes disregarded as the introduction of potential new quantum (gravity) effects could change the scenario drastically, avoiding the formation of singularities. See also \cite{Feng:2023pfq}.  

It is well known that an invariant associated with the measurement of tidal forces is the Kretschmann scalar \cite{Bena:2020iyw}. In a four dimensional vacuum spherically symmetric space, this scalar is proportional to \( K \sim \bar{M}^2/r^6 \), leading to infinite tidal forces at the origin. Gravitational field tension in the spherically symmetric case is characterized by the curvature term given by the square root of the Kretschmann scalar of the vacuum theory, \( F \sim \sqrt{K_{\text{Schw}}} \sim \frac{\bar{M}}{r^3} \) in $4D$. This correlation is logical, as the spacetime tension should increase with the mass of the vacuum source \cite{Alencar:2023wyf}. In this sense, Dymnikova's energy density \cite{Dymnikova:1992ux} can be interpreted as:

\begin{equation} \label{probability} 
{\Huge \rho}_{\mbox{\small--Dymnikova}} \sim \exp \left ( - \frac{F_c}{F}\right)       \end{equation}                             
being $F_c,a$ constants. It is worth mentioning that in the four-dimensional case, this energy density behaves as follows:
\begin{equation}
 {\Huge \rho}_{\mbox{\small--Dymnikova}}   \sim \exp \left ( - \frac{r^3}{a^3 }  \right)
\end{equation}

Dymnikova proposes a model that encodes the gravitational information of the spherically symmetric vacuum solution in such a way that near the central singularity—where the gravitational tension and tidal forces of the vacuum black hole become infinite—the energy density in the Dymnikova model (DyM) remains finite. This premise will be crucial for the model we examine in this work. This finite density in the DyM behaves as a positive cosmological constant near the origin, resulting in a regular black hole solution with a de Sitter core.  It is also worth noting that the DyM has garnered significant attention in recent years for addressing various issues in physics \cite{Paul:2023pqn,Konoplya:2024kih,Estrada:2023pny,Alencar:2023wyf,Estrada:2019qsu,Bueno:2024dgm}.

Reference \cite{DymnikovaS1996} suggests an interesting speculative analogy between the DyM and the quantum Schwinger effect: the density of the DyM attained during collapse could approach the Planck scale or possibly the GUT scale, depending on the nature of the fields in the energy-momentum tensor. Thus, the energy-momentum tensor encapsulates the effects of these fields, relating them to the gravitational field tension.  In this framework, reference \cite{Ansoldi:2008jw} considers the case where the gravitational source is of electric origin, \( E \sim F \). Recently, references \cite{Alencar:2023wyf,Estrada:2019qsu} explored this analogy for regular black holes and wormholes under the influence of quantum GUP. However, despite what has been discussed in this paragraph, a deeper model for this quantum analogy would require a more thorough investigation, which is beyond the scope of this work. It is worth mentioning that in a very different framework is also associated the value of an electric field in the Schwinger effect with the value of the Kretschmann scalar in references \cite{Wondrak:2023zdi,Chernodub:2023pwf}. 

On the other hand, it is well-known that several branches of theoretical physics predict the existence of extra dimensions. Even though several experiments have tried to test this idea, this is yet to be observed. Consequently, any theory incorporating extra dimensions must align with General Relativity in four dimensions or with one of its generalizations. Among these theories is Lovelock gravity. The Lagrangian of Lovelock gravity includes higher curvature terms as corrections to the Einstein-Hilbert action \cite{Lovelock:1971yv}. Furthermore, Lovelock's theories adhere to the fundamental principles of General Relativity; for example, their equations of motion are of second order. It is important to mention that the specific case of Lovelock gravity, known as Einstein-Gauss-Bonnet theory, has garnered attention in recent years for its applications in inflationary theories and has been compared with the results from GW170817 \cite{Odintsov:2020zkl,Oikonomou:2021kql}. 

A special case of Lovelock theories is Pure Lovelock theory (PL) \cite{Cai:2006pq}. As will be discussed below, this theory considers a single term in the Lagrangian. Pure Lovelock theory has drawn attention in recent years for various problems in physics \cite{Singha:2023lum,Paithankar:2023ofw,Shaymatov:2020byu}. See also \cite{Chakraborty:2016qbw,Chakraborty:2015kva} . As we will see below, the value of the Kretschmann scalar for the vacuum solution depends on the power \( n \) of the curvature. This makes it interesting to study the behavior of a non-vacuum model that encodes the gravitational information of the empty geometry in an analogous way to the Dymnikova model.

Pure Lovelock is a theory characterized by specific properties that distinguish it from general Lovelock theory and all other higher derivative theories:  vacuum solutions of the motion equations in Pure Lovelock theory are doubly degenerate for even \( n \), a situation that does not occur in General Relativity where \( n = 1 \). This also raises an intriguing question about what these degenerate solutions represent in the non-vacuum case and whether they allow for transverse geometries that differ from spherical symmetry. Additionally, it is well known that General Relativity has a non-trivial vacuum solutions, where the gravitational potential of the form \( -\bar{M}/r^{d-2n-1} \) does not depend radially on its denominator when \( d = 3 \) (i.e., \( d = 2n + 1 \) where \( n = 1 \)). An interesting feature is that pure Lovelock theory retains this property for \( d = 2n + 1 \) with \( n > 1 \). Related to this, references \cite{Dadhich:2015lra,Dadhich:2012cv,Camanho:2015hea} discuss a universal property of pure Lovelock theory, highlighting its kinematic nature in critical odd dimensions \( d = 2N + 1 \). This property raises the question of what the physical interpretation would be in the non-vacuum case, i.e., when \(d = 2n + 1\) in the presence of high curvature terms. Another remarkable property of Lovelock theory is the existence of bound orbits in higher dimensions \cite{Dadhich:2013moa}.

In this work, we present a regular black hole model for Pure Lovelock gravity, based on the previously described idea: a model of energy density that encodes the gravitational information of the vacuum case, ensuring that tidal forces and energy density remain finite at the radial origin. We will analyze what happens for each value of the power of the Riemann tensor in each theory with $n > 1$. In the case where degenerate solutions exist, we will investigate whether there are black hole solutions with transverse geometries different from the usual spherically symmetric ones. We will also examine the physical interpretation of the solutions obtained for the cases $d = 2n+1$ and $d > 2n+1$. As we will discuss later, we will interpret the horizon structure, identifying scenarios with the presence and absence of inner and cosmological horizons, in addition to the event horizon. For all the cases studied, we will investigate the temperature and radial evolution, which will provide insights into the evaporation process and help determine under what conditions the final stage of this process would correspond to a remnant.

On the other hand, in recent years, there has been growing interest in determining whether the presence of an inner horizon is inherently unstable. For example, reference \cite{Carballo2023} asserts that instability in the cores of regular black holes is inevitable because mass inflation instability is crucial for regular black holes with astrophysical significance. However, this remains an unresolved issue from a theoretical standpoint. Conversely, reference \cite{Bonanno:2022jjp} argues that semiclassical effects due to Hawking radiation might mitigate the instability associated with the inner horizon, potentially stabilizing the existence of a de Sitter core. Therefore, in this work, we will also test whether black holes with only an event horizon and no inner horizon can exist in non-spherical transverse geometries for $d=2n+1$ and $d>2n+1$.

\section{A brief review about Lovelock gravity and the Pure Lovelock case}

The Lovelock Lagrangian \cite{Lovelock:1971yv} is :

\begin{equation}\label{LovelockLagrangian}
L  = \sqrt{-g} \sum_{n=0}^N \alpha_n L_n,
\end{equation}
where \( N = \frac{d}{2} - 1 \) for \( d \) even and \( N = \frac{d-1}{2} \) for \( d \) odd, and \( \alpha_n \) are arbitrary coupling constants. \( L_n \) is a topological density defined as:
\begin{equation}
 L_n = \frac{1}{2^n} \delta^{\mu_1 \nu_1 ...\mu_n \nu_n}_{\alpha_1 \beta_1 ... \alpha_n \beta_n} \displaystyle \Pi^n_{r=1} R^{\alpha_r \beta_r}_{\mu_r \nu_r},
\end{equation}
where \( R^{\alpha \beta}_{\mu \nu} \) is an \( n \)-order generalization of the Riemann tensor for the Lovelock theory, and:
\begin{equation}
 \delta^{\mu_1 \nu_1 ...\mu_n \nu_n}_{\alpha_1 \beta_1 ... \alpha_n \beta_n} = \frac{1}{n!} \delta^{\mu_1}_{\left[\alpha_1\right.} \delta^{\nu_1}_{\beta_1} ... \delta^{\mu_n}_{\alpha_n} \delta^{\nu_n}_{\left.\beta_n \right]}   
\end{equation}
is the  generalized Kronecker delta.

It is important to emphasize that the terms $L_0$, $L_1$, and $L_2$ are proportional to the cosmological constant, the Ricci scalar, and the Gauss-Bonnet Lagrangian, respectively. The corresponding equation of motion is given by:
\begin{equation}
 \sum ^N_{n=0} \alpha_n \mathcal{G}^{(n)}_{AB} =   T_{AB},
\end{equation}
where $\mathcal{G}^{(n)}_{AB}$ represents an $n$-order generalization of the Einstein tensor, influenced by the topological density $L_n$. For example, $\mathcal{G}^{(1)}_{AB}$ corresponds to the Einstein tensor related to the Ricci scalar (with Einstein-Hilbert theory as a specific case of Lovelock theory), and $\mathcal{G}^{(2)}_{AB}$ corresponds to the Lanczos tensor associated with the Gauss-Bonnet Lagrangian.

\subsection{Pure Lovelock case} \label{PureLovelock}

Pure Lovelock theory involves only a single fixed value of $n$ (with $n \geq 1$), without summing over lower orders. In some cases, it is considered as a single value of $n \geq 1$ plus the $n=0$ term, i.e., $L = L_0 + L_n$, as shown in references \cite{Aros:2019quj, Toledo:2019mlz, Toledo:2019szg}. For simplicity, in this work, we consider only the $L_n$ term without the $L_0$ term, as illustrated in references \cite{Dadhich:2015rea, Dadhich:2016wtb}. Thus, the Lagrangian is:
\begin{equation}
 L = \sqrt{-g} \alpha  L_n= \sqrt{-g} \alpha_n \frac{1}{2^n} \delta^{\mu_1 \nu_1 ...\mu_n \nu_n}_{\alpha_1 \beta_1 ... \alpha_n \beta_n} \displaystyle \Pi^n_{r=1} R^{\alpha_r \beta_r}_{\mu_r \nu_r},
\end{equation}

The equations of motion are given by:

\begin{equation} \label{eqmovimiento}
 \mathcal{G}^{(n)}_{AB} = T_{AB} , 
\end{equation}
where
\begin{equation}
    (\mathcal{G}^{(n)})_{B}^A=-\frac{1}{2^{n+1}}\,\delta_{B A_{1}...A_{2n}}^{A
B_{1}...B_{2n}}\,R_{B_{1} B_{2}}^{A_{1} A_{2}}\cdots R_{B
_{2n-1} B_{2n}}^{A_{2n-1} A_{2n}}\,.
\end{equation}
And the coupling constants were set to unity, consistent with references \cite{Dadhich:2015rea, Dadhich:2016wtb}.

\section{Our Higher Dimensional model}

The vacuum solution in Pure Lovelock gravity can be found in reference \cite{Cai:2006pq}, and is given by:
\begin{equation} \label{VacioCai}
    f(r)= \gamma- \left ( \dfrac{2\bar{M}}{r^{d-2n-1}}  \right )^{1/n}
\end{equation}
where, as we will see below, $\gamma$ is an integration constant related to the geometry of the non-transversal section \cite{Aros:2000ij}. On the other hand, the Kretschmann invariant is given by
\begin{equation}
  K=  \left(f''(r)\right)^2 + \frac{2(d - 2)}{r^2}\left(f'(r)\right)^2 + \frac{2(d - 2)(d - 3)}{r^4}\left(\gamma - f(r)\right)^2
\end{equation}
Evaluating solution \eqref{VacioCai} in the previous equation:
\begin{equation}
    K= C \frac{\bar{M}^{2/n}}{r^{(2/n)(d-1)}}
\end{equation}
where
\begin{eqnarray*}
C &= & \frac{ 4^{\frac{1}{n}}}{n^{4}} \left(2n^4d(d - 3) + (8d - 16)n^4 + (-8d^2 + 12d - 4)n^3 \right. \\
&+& \left. (2d^3 + 5d^2 - 16d + 9)n^2 - 6(d - 1)^3n + (d - 1)^4\right)  
\end{eqnarray*}
Following the previously described idea and motivated by Dymnikova's model, we propose a model where the energy density encodes the gravitational information of the vacuum solution with constant transversal curvature. In our model, near the central singularity—where the gravitational tension and tidal forces of the vacuum black hole become infinite—the energy density assumes a finite value.

As previously mentioned, the gravitational tension is associated with the square root of the Kretschmann scalar of the vacuum solution, which, in this case, corresponds to pure Lovelock gravity:

\begin{equation}
    F \sim \sqrt{K} \sim \frac{\bar{M}^{1/n}}{r^{(d-1)/n}}
\end{equation}

Thus, analogous to equation \eqref{probability}, we will define the energy density as:

\begin{equation} \label{DensidadDeEnergia}
\rho = A \exp \left ( - \frac{r^{(d-1)/n}}{a^{(d-1)/n}} \right )
\end{equation}
where $a$ is a constant and where, for simplicity, the constant $A$ has been adjusted to:
\begin{equation} \label{ADensidadDeEnergia}
    A= \frac{d-2}{n} \frac{M}{a^{d-1}/(d-1)}
\end{equation}

\section{Our regular black hole solution in Pure Lovelock gravity}
In this work, we study the static $d$-dimensional metric, which is given by:
\begin{equation} \label{metrica}
    ds^2=-\mu(r)dt^2+\mu(r)^{-1} dr^2+r^2 d\Sigma_\gamma. 
\end{equation}

The constant $\gamma$ can be normalized to $\pm 1$, $0$ by appropriately rescaling. Thus, the local geometry of $\Sigma_\gamma$ is a sphere, a plane, or a hyperboloid \cite{Aros:2000ij}:
\[
\Sigma_\gamma \text{ locally } =
\begin{cases}
S^{d-2}& \text{for } \gamma = 1 \\
T^{d-2} & \text{for } \gamma = 0 \\
H^{d-2}/G_{H} & \text{for } \gamma = -1.
\end{cases}
\]
Here $\gamma$ stands for the value of the local (constant Riemannian) curvature. $G_{H}$ could be any element of $SO(1,d-3)$ which acts freely on $H^{d-2}$ and such that the quotient be compact. In the same fashion for $\mathbb{R}^{d-2}$, $T^{d-2}$ is the $d-2$-torus, i.e., $(S^1)^{d-2}$. 

The energy-momentum tensor corresponds to a neutral perfect fluid:
\begin{equation}
T^A_B=\mbox{diag}(-\rho,p_r,p_\theta,p_\theta,...),    
\end{equation}
On the one hand, it is well known that this form of the metric imposes the condition \(\rho = -p_r\), which implies that the \((t,t)\) and \((r,r)\) components of the equations of motion have the same structure, given by:
\begin{equation} \label{ttcomponente}
 \frac{d}{dr} \Big (r^{d-2n-1} \left( \gamma- \mu \right) ^n    \Big ) = \frac{2}{d-2} r^{d-2} \rho ,
\end{equation}

On the other hand, due to transversal symmetry, we have \(p_\theta = p_\phi = p_t = \ldots\) for all the \((d-2)\) angular coordinates. Thus, using the aforementioned condition \(\rho = -p_r\), the conservation law \(T^{AB}_{;B} = 0\) gives:
\begin{equation} \label{conservacion1}
p_t = -\frac{r}{d-2} \rho' - \rho 
\end{equation}

Defining the mass function as:
\begin{equation} \label{FuncionMasaIntegral}
    m(r)= \frac{1}{d-2}  \int \rho r^{d-2} dr
\end{equation}

By choosing $M \cdot(n-1)!$ as integration constant, the mass function is given by

\begin{equation} \label{FuncionMasaGamma}
m(r) = M \left ( (n-1)! - \Gamma \left [ n \, , \, \frac{r^{(d-1)/n}}{a^{(d-1)/n}} \right ]  \right ) = \frac{M}{n} {}_{1}F_{1}\left(n,n+1,-\left(\frac{r}{a}\right)^{\frac{d-1}{n}}\right) \left(\frac{r}{a}\right)^{d-1}
\end{equation}
where $\Gamma$ is the Gamma function and where ${}_1F_{1}$ is hyperbolic confluent function. As expected, provided $r\rightarrow \infty$ describe a proper region of the space, 
\begin{equation}
    \lim_{r\rightarrow \infty} m(r) = (n-1)! M 
\end{equation}

On the other hand, for small values of $r$, such that \( r \ll a \), it is satisfied 
\begin{equation}
    m(r)|_{r \ll a} \approx \frac{M}{a^{d-1} \cdot n} r^{d-1},
\end{equation}

The solution of equation \eqref{ttcomponente} for $n=$odd is given by

\begin{equation}
    \mu(r)=\gamma- \left ( \frac{2 m(r)}{r^{d-2n-1}}  \right)^{1/n}
\end{equation}

Thus, {\bf for odd \( n \)}, we will focus only on the solution where \( \gamma = 1 \), whose transversal section corresponds to a sphere, since, as we will see below, this is the only case that has horizons.

\begin{equation} \label{SolucionGamma1}
    \mu(r)_{\gamma=1}=1- \left ( \frac{2 m(r)}{r^{d-2n-1}}  \right)^{1/n}
\end{equation}

However, for {\bf even \( n \)}, the equation \eqref{ttcomponente} has the following solutions:

\begin{equation}
    \mu(r)=\gamma \pm \left ( \frac{2 m(r)}{r^{d-2n-1}}  \right)^{1/n}
\end{equation}

For even $n$, we will be interested in the following cases, which, as we will see below, possess horizons:

\begin{itemize}
    \item One of them corresponds to the negative branch of the previous equation with \( \gamma = 1 \), which corresponds to Equation \eqref{SolucionGamma1}.
    \item The other corresponds to the positive branch of the previous equation with \( \gamma = -1 \), which corresponds to:
\begin{equation} \label{SolucionGamma-1}
    \mu(r)_{\gamma=-1}=-1+ \left ( \frac{2 m(r)}{r^{d-2n-1}}  \right)^{1/n}
\end{equation}
\end{itemize}

From the above, we can note the following characteristics:

\begin{itemize}
    \item Solution \eqref{SolucionGamma1} represents a black hole solution with a cross-section that corresponds to an \( S_{d-2} \) sphere.
    \item Solution \eqref{SolucionGamma-1} represents a black hole solution with a cross-section that corresponds to an \( H_{d-2} \) hyperboloid.
    \item Both solutions are asymptotically flat
    \begin{equation} \label{ConvergenciadS}
        \displaystyle \lim_{r \gg a} \mu(r)_{\gamma=1} = 1 - \left(\frac{2(n-1)!M}{r^{d-2n-1}}\right)^\frac{1}{n},
    \end{equation}
    \begin{equation} \label{ConvergenciaAdS}
        \displaystyle \lim_{r \gg a} \mu(r)_{\gamma=-1} = -1 + \left(\frac{2(n-1)!M}{r^{d-2n-1}}\right)^\frac{1}{n} 
    \end{equation}
It can be noticed that any information of $a$ becomes undetectable in the asymptotic flat region. 
\item It is worth mentioning that the solution with $\gamma=1$ behaves near the origin as
\begin{equation} \label{NucleoDeSiter}
   \lambda(r)= \mu (r)_{\gamma=1} \Big|_{r \ll a}  1- \left ( \frac{2M}{a^{d-1} \cdot n} \right )^{\frac{1}{n}} r^2.
\end{equation}
Therefore, this later solution possesses a de Sitter core, representing a regular black hole. Thus, at small scales, the behavior of the energy density results in a de Sitter core instead of a central singularity.
\item Also we can check that the solution with $\gamma=-1$ and even $n$ behaves near the origin as
\begin{equation} \label{NucleoAdS}
   \mu (r)_{\gamma=-1} \Big|_{r \ll a}  -1+ \left ( \frac{2M}{a^{d-1} \cdot n} \right )^{\frac{1}{n}} r^2.
\end{equation} 
Therefore, this later solution possesses an Anti de Sitter core, representing a regular black hole.
\end{itemize}

On the other hand, it is worth noting that both cases, with \( \gamma = 1 \) and \( \gamma = -1 \), have a finite value of the Kretschmann scalar, so both solutions correspond to regular black holes.

\begin{equation}
    K \approx d(d-1) \left ( \frac{24^{1/2}M}{n a^{d-1}} \right )^{2/n}
\end{equation}

\section{Horizon structure and physical interpretation}

Given from of the metric (\ref{metrica}), \eqref{FuncionMasaGamma}, for any value $r=r_s$ such that $\mu(r_s)=0$ the space presents a Killing horizon.  Now, for both cases, equations \eqref{SolucionGamma1} and \eqref{SolucionGamma-1}, the zeros of $\mu(r)$ can be obtained from the condition
\begin{equation}\label{rplusSol}
   M(r_s)= r_s^{(-1 + d - 2 n)} \bigg/ \left( 2 \left( (-1 + n)! - \Gamma\left[n, a^{\frac{1 - d}{n}} r_s^{\frac{-1 + d}{n}}\right] \right) \right)
\end{equation}
where \( M(r_s) \) corresponds to the mass parameter. In what follows, we will analyze separately the branch with \(\gamma = 1\) (with \(n\) even and odd) and the branch with \(\gamma = -1\) (with \(n\) even).

For the analysis below, it will be useful to express this equation in terms of \( r_s/a \) with \( r = r_s \):

\begin{equation} \label{MasaHipergeometrica}
\frac{M[r_s/a]}{a^{d-2n-1}} = \tilde{M}= \frac{n}{2} \left(\frac{a}{r_s}\right)^{2n}\frac{1}{{}_1 F_{1}\left(n,n+1,-\left(\frac{r_s}{a}\right)^{\frac{d-1}{n}}\right)}.
\end{equation}
which resembles a parabolic curve for $d-2n-1>0$. As we will see below, for $d-2n-1>0$ this implies that the presence of a minimal mass at a value \( r_s = r_* = r_+ = r_- \) (or \( r_s = r_* = r_+ = r_{++} \)) with \( M(r_*) = M_0 > 0 \), where $r_-$, $r_+$, and $r_{++}$ represent the inner, event, and cosmological horizons, respectively. This is the case where the temperature of the system vanishes. See below.

\subsection{branch with \(\gamma = 1\) for both even and odd \(n\)}

\subsubsection{\bf $d=2n+1$}

\begin{figure}[h]
    \centering
    \includegraphics[scale=.4]{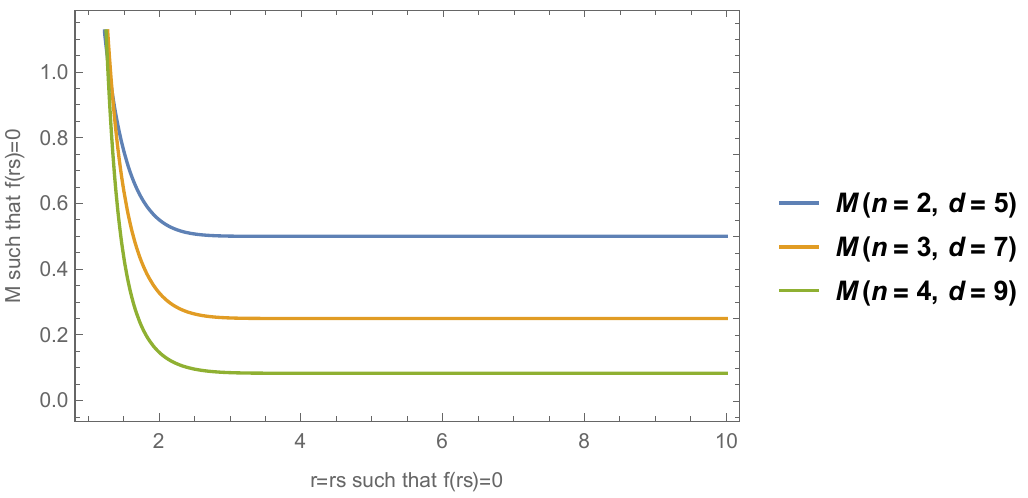}
    \includegraphics[scale=.4]{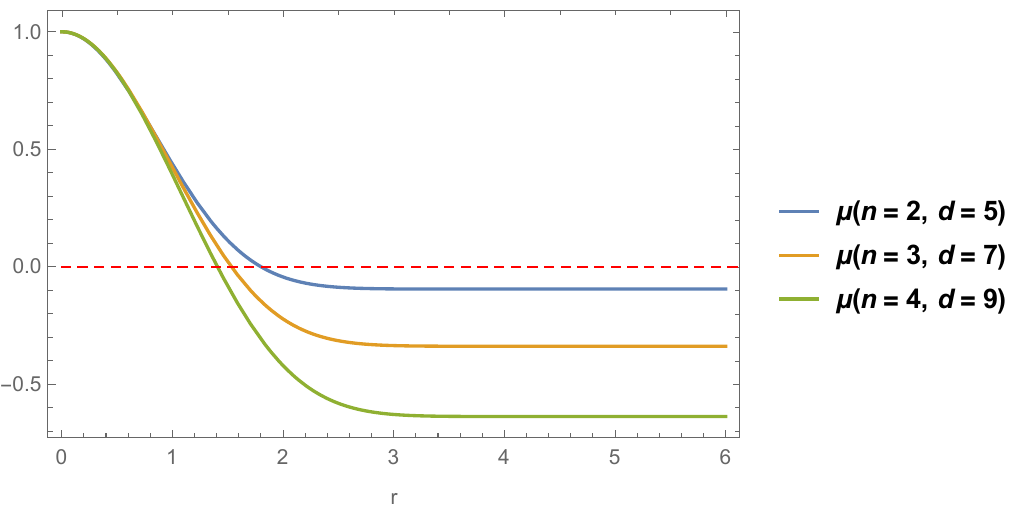}
    \caption{Left side: Parameter of $M$ for $a=1$. Right side: Function $\mu$ for $a=1$, $M=0.6$.}
    \label{M1d=2n+1}
\end{figure}

After a straightforward analysis, as shown in Figure \ref{M1d=2n+1}, it can be observed that when \(d = 2n + 1\), for each value of \(r_s\), there is a unique value for the parameter \(M\). Thus, there is always a single solution, namely \(r_s = r_{++}\), with \(f(r) > 0\) for \(r < r_{++}\) and \(f(r) < 0\) for \(r > r_{++}\). On the other hand, if we denote by $M_*$ the asymptotic value of the vertical axis in the figure, we can observe that a condition for the existence of horizons is that $M > M_*$. Consequently, the geometry can be interpreted as a cosmology described in the static region, with \(r_{++}\) defining the cosmological horizon. It should be noted that in this case, the parameter \(M\) cannot be interpreted as a Noether charge or a Hamiltonian quantity. Therefore, \(M\) does not correspond to the mass of the solution.

\subsubsection{\bf $d>2n+1$}

\begin{figure}[h]
    \centering
    \includegraphics[scale=.4]{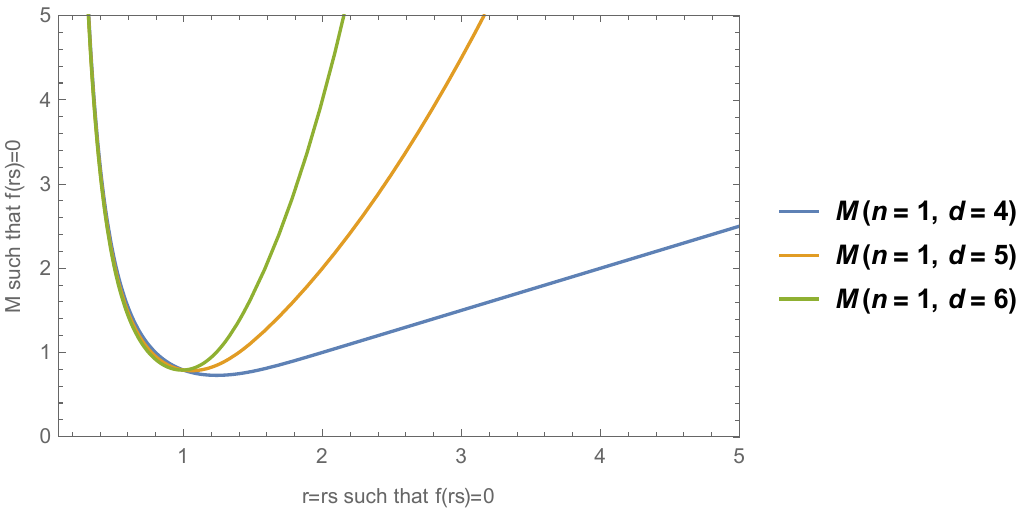}
    \includegraphics[scale=.4]{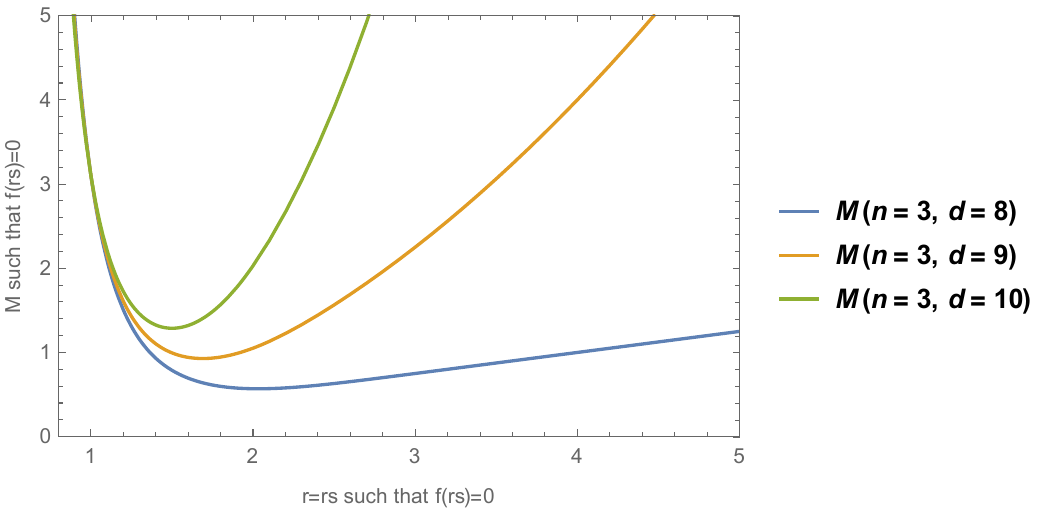}
    \caption{Parameter of $M$ for $a=1$ and odd $n$.}
    \label{MImpar}
\end{figure}

\begin{figure}[h]
    \centering
    \includegraphics[scale=.4]{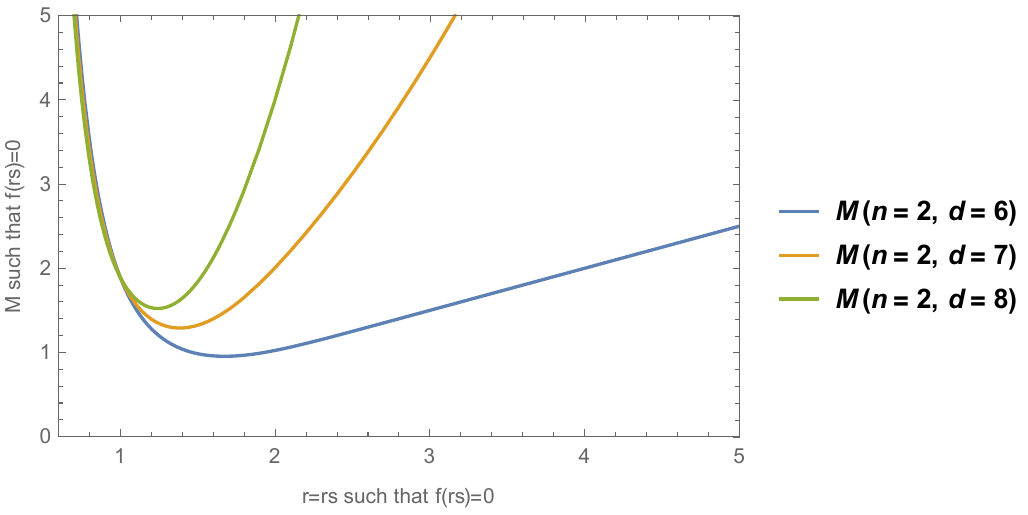}
    \includegraphics[scale=.4]{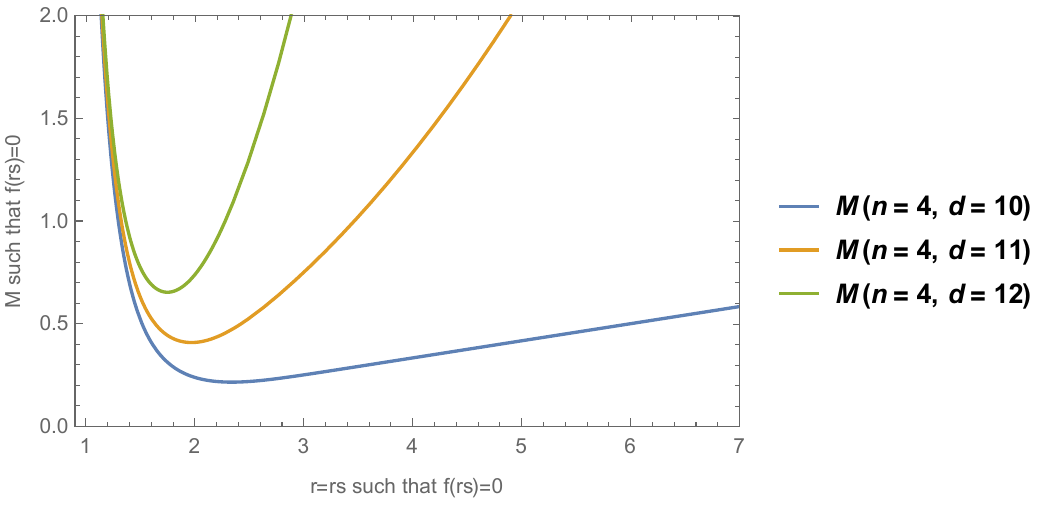}
    \caption{Parameter of $M$ for $a=1$ and even $n$.}
    \label{Mpar}
\end{figure}

In Figures \ref{MImpar} and \ref{Mpar}, we observe the behavior of the parameter \( M \) for different values of \( n \) and \( d \). It is evident that, in all cases, \( M \) attains a minimum value \( M(r_*) \) when the inner and event horizons coincide, i.e., \( r_s = r_* = r_- = r_+ \). The values of \( r_s \) located to the left of \( r_* \) correspond to the possible values of the inner horizon. Conversely, the values of \( r_s \) located to the right of \( r_* \) correspond to the possible values of the event horizon. 

\begin{figure}[h]
    \centering
    \includegraphics[scale=.4]{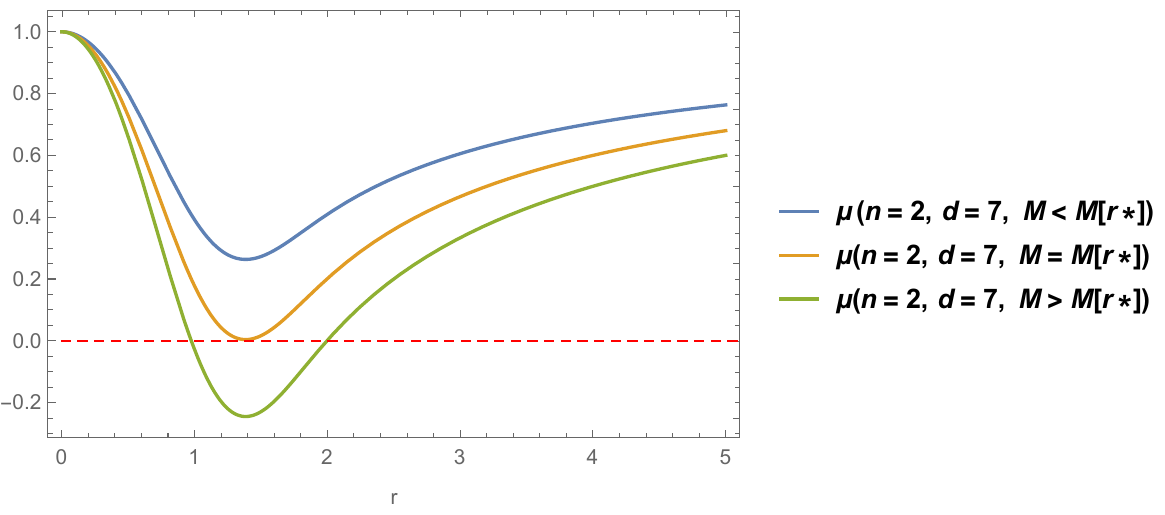}
    \includegraphics[scale=.4]{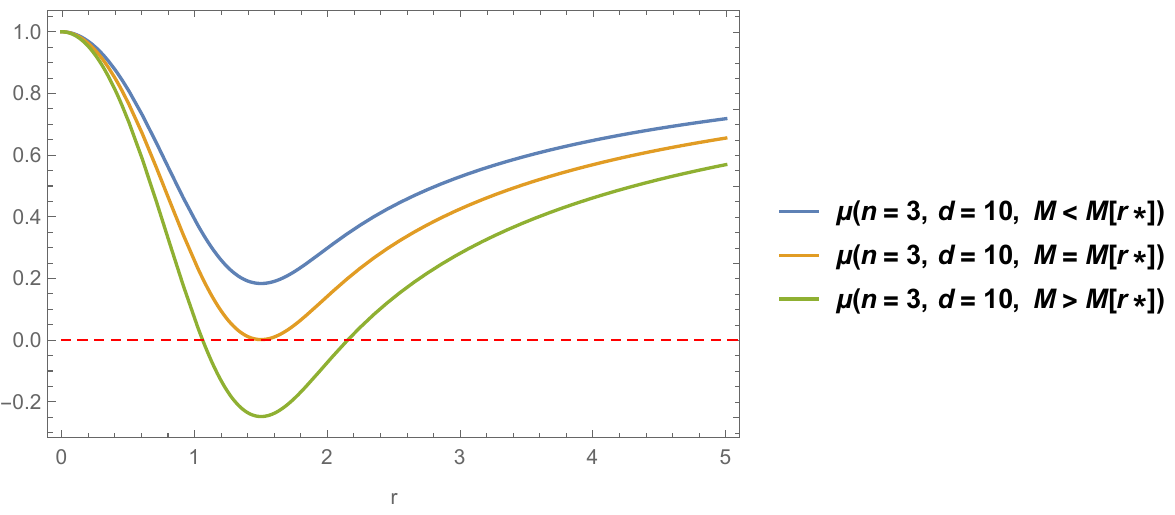}
    \caption{Left side: Function $\mu$ for $a=1$, $n=2$, $d=7$, $M(r_*)=1.28$. Right side: Function $\mu$ for $a=1$, $n=3$, $d=10$, $M(r_*)=1.28$.}
    \label{f1}
\end{figure}

As observed in the examples displayed in Figure \ref{f1}, where Equation \eqref{SolucionGamma1} is plotted on the vertical axis, the structure of the zeros can be categorized into three distinct cases. It is straightforward to verify that this behavior is similar for other values of \( n \) and \( d \).

\begin{itemize}
    \item For $M>M(r_*)$, $\mu(r)=0$ $\mu(r)=0$ has two zeros, $r_+ > r_-$, and the geometry corresponds to a regular black hole with an external and internal horizon. In this case, $r_+$ determines the event horizon. In this case the parameter $M$ represents the mass/energy of the solution as expected above. One very interesting feature of Eq.\eqref{SolucionGamma1} is how fast the regular solution converges into the non-regular solution in  \cite{Cai:2006pq}. This can be glimpsed in Fig \ref{convergency} where the respective $\mu(r)$ functions are displayed as functions $r/a$ for the same $M/a^{d-2n-1}$. In fact, it seems as the two functions would match between $r_{-}<r<r_{+}$. In reality, the separation between the two functions falls below $\epsilon \ll 1$ rapidly. This implies that is basically impossible to separate both solutions outside of the outermost horizon.
\begin{figure}[h]
  \centering
  \includegraphics[width=3in]{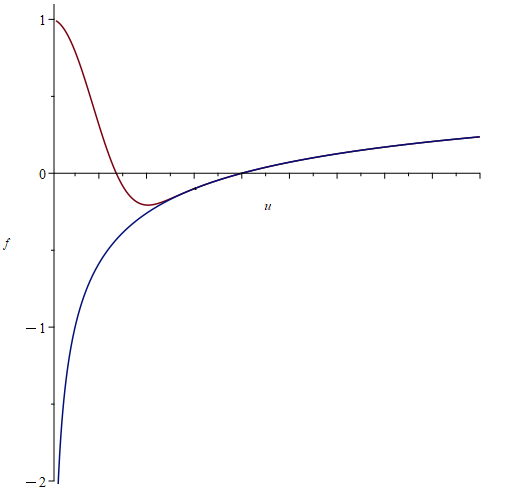}
  \caption{Comparison between regular solution with two horizons and the standard black hole solution}\label{convergency}
\end{figure}
    \item For $M=M(r_*)$, has one (double) zero, $r_+ = r_-=r_s$, and the geometry corresponds to a zero-temperature regular black hole. See below. As before, the $M$ parameter corresponds to the mass of the solution. This solution has no standard black hole counterpart. This case is also very interesting as the regular solution and the standard solution in \cite{Cai:2006pq} have very distinct thermodynamical properties, see below. For instance the regular solution would not have a Hawking radiation.
    \item For $M<M(r_*)$, $\mu(r)=0$ has no (real) solutions. In this case, the space is a regular space with no horizons. It must be emphasized that the lack of horizons, within this family of geometries, does not imply the presence of singularities. As before, the energy of the solution is given by $M$. In a matter of speaking this solution can be understood as star-like solution. In must be emphasized that the counter part of this case in \cite{Cai:2006pq} is a black hole solution.    
\end{itemize}

It must be stressed that even though the concept of mass of the solutions is always well-defined for $d>2n+1$, and the geometry is non-singular, only for the case $r_+>r_-$ is possible construct a thermodynamics. In this case the mass $M$ can be written as a function of $r_+$ or $r_-$ and it is given by equation \eqref{rplusSol}.

\subsection{branch with \(\gamma = -1\) for even $n$}

\subsubsection{\bf $d=2n+1$} \label{Rama-1d2n+1}
On one hand, we can observe that for even values \( n = 2 \) and \( n = 4 \), the behavior of the mass parameter is the same on the left side of Figure \ref{M1d=2n+1}. In other words, there can be at most one horizon for a given value of \( M \), and a necessary condition for the existence of such a horizon is that \( M > M(r_*) \).

\begin{figure}[h]
    \centering
    \includegraphics[scale=.4]{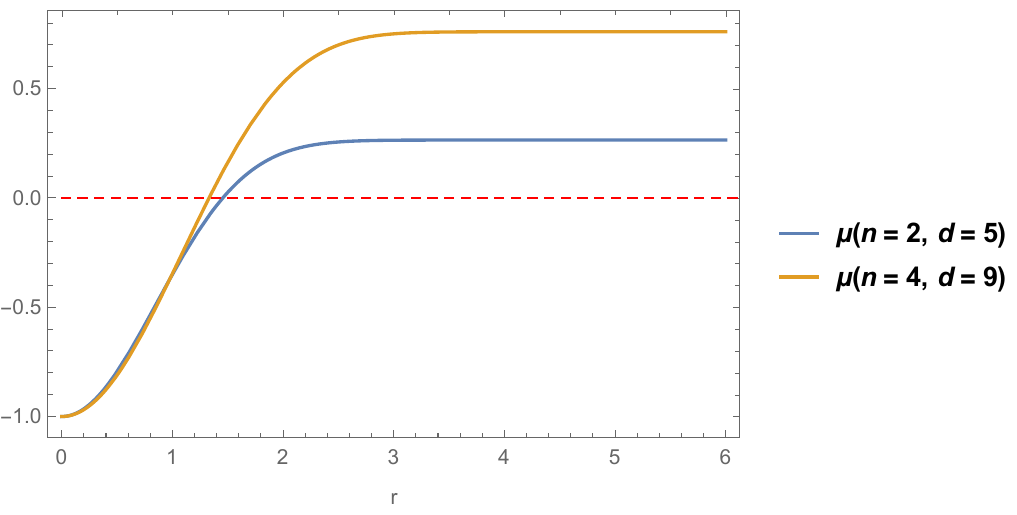}
       \caption{Function $\mu$ for $a=1$, $M=0.6$.}
    \label{f-1d=2n+1}
\end{figure}

Thus, as observed in Figure \ref{f-1d=2n+1}, the signature outside the horizon \( r_s \) corresponds to that of a black hole. Therefore, this solution represents a regular black hole, and \( r_s = r_+ \) corresponds to the event horizon. Consequently, the parameter \( M \) represents the mass of the black hole.

This is an interesting result, as it represent to a regular black hole solution without the presence of an inner horizon or a de Sitter core for $n=2,d=5$ and $n=4,d=9$. In this regard, it is important to mention that some authors associate the presence of an inner horizon with a point where predictability fails \cite{Ovalle:2023vvu}. It has also been discussed that instabilities at the inner horizon can lead to mass inflation and disrupt fundamental physics \cite{Poisson:1990eh,Brown:2011tv}. Recently, reference \cite{Carballo2023} asserts that instability in the cores of regular black holes is inevitable because mass inflation instability is crucial for regular black holes with astrophysical significance. However, this remains an unresolved physical problem from a theoretical perspective. Regarding this issue, see reference \cite{Bonanno:2022jjp}, where it is argued that semiclassical effects due to Hawking radiation might alleviate the instability associated with the inner horizon, thereby making the existence of a de Sitter core stable.

\subsubsection{\bf $d>2n+1$}
\begin{figure}[h]
    \centering
    \includegraphics[scale=.4]{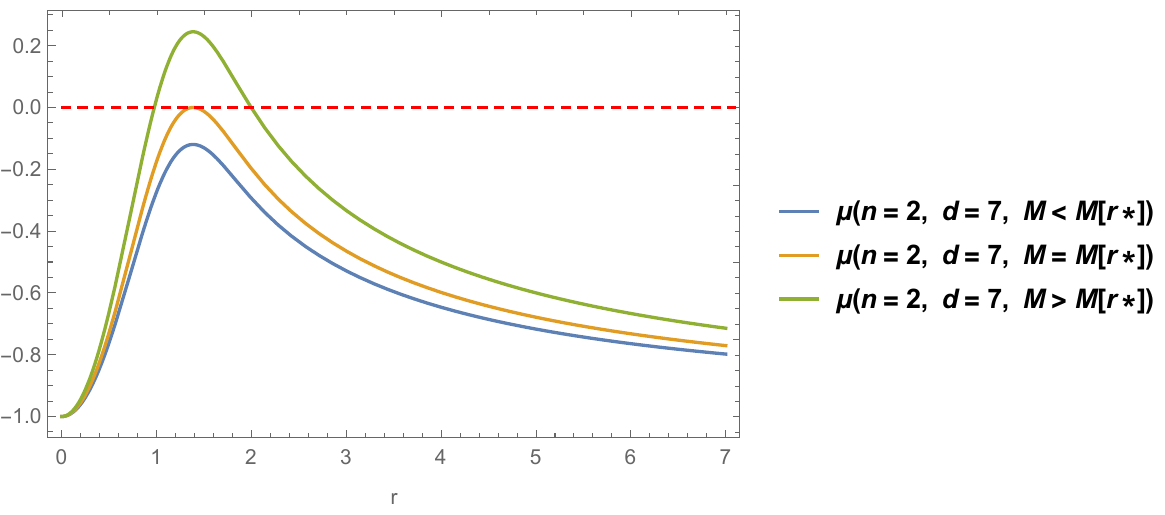}
    \includegraphics[scale=.4]{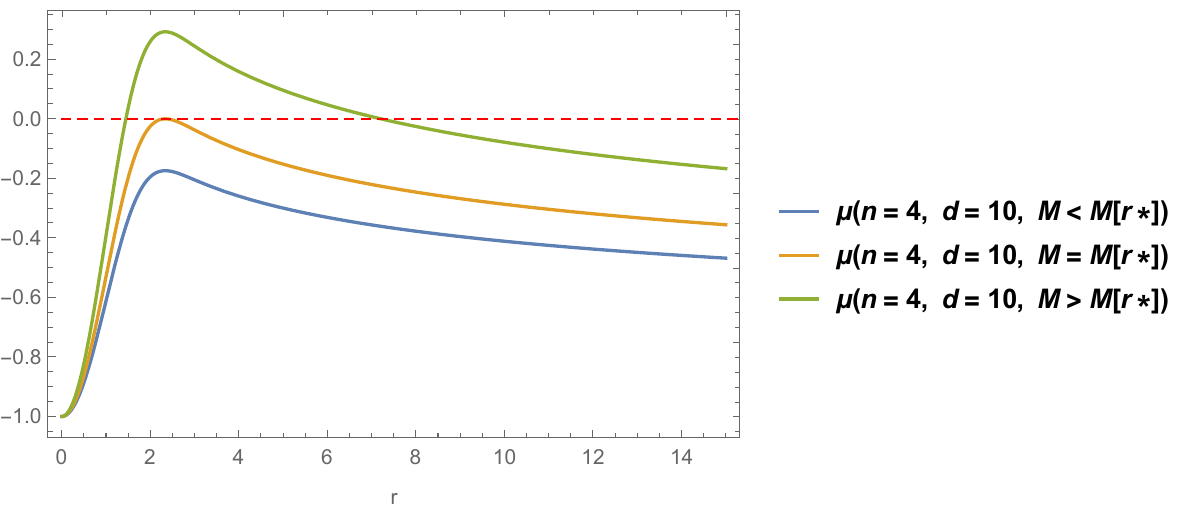}
    \caption{Left side: Function $\mu$ for $a=1$, $n=2$, $d=7$, $M(r_*)=1.29$. Right side: Function $\mu$ for $a=1$, $n=4$, $d=10$, $M(r_*)=0.215$.}
    \label{f-1}
\end{figure}

On one hand, we can observe that for even values \( n = 2 \) and \( n = 4 \), the behavior of the mass parameter is consistent, as shown in Figure \ref{Mpar}. Thus, as we can see in figure \ref{f-1}, we can identify the following cases:

\begin{itemize}
    \item For \( M < M(r_*) \), the equation \(\mu(r) = 0\) has no real solutions. In this case, the spacetime, similar to the branch with \(\gamma = 1\), is also regular and free of horizons. Additionally, the absence of horizons in this family of geometries does not imply the presence of singularities. As before, the energy of the solution is given by \( M \).
    \item For \( M > M(r_*) \), the equation \(\mu(r) = 0\) has two zeros, \( r_+ < r_{++} \), and the geometry corresponds to a regular black hole with both an event horizon and a cosmological horizon. In this case, \( r_{++} \) defines the outermost horizon. The presence of the cosmological horizon prevents the computation of energy using Noether's charge; thus, the parameter \( M \) is not directly associated with the mass. However, using thermodynamic arguments, a local definition of mass variation can be established, which depends on \( M \) \cite{Aros:2008ef}. Thus, a remarkable feature of this solution is the presence of both an event horizon and a cosmological horizon without the presence of a positive cosmological constant, as seen in the Schwarzschild-de Sitter case.

Furthermore, similar to the case where \( d = 2n + 1 \), sub section \ref{Rama-1d2n+1}, this solution represents a regular black hole without an internal horizon or a de Sitter core. As mentioned earlier, some authors argue that the de Sitter core might be unstable due to mass inflation \cite{Carballo2023}, while others argue the contrary \cite{Bonanno:2022jjp}.
    \item For \( M = M(r_*) \). This point represents where the event horizon and the cosmological horizon coincide. This situation is analogous to what occurs in the dS-Schwarzschild black hole solution when studying the thermodynamic equilibrium \cite{Ginsparg:1982rs}. The latter reference describes that the Nariai solution \cite{Nariai1999} is obtained when the event horizon of the dS-Schwarzschild black hole approaches the cosmological horizon. 
\end{itemize}

Thus, it is important to mention that for $d=2n+1$, the ordered pairs $(r_s, M(r_s))$ correspond to a cosmological horizon for the branch with $\gamma=1$ and to an event horizon for the branch with $\gamma=-1$ with $n$ even. 

For $d>2n+1$, there are two values of $r_s$ such that $M(r_{s1})=M(r_{s2})$, with $r_{s1} < r_{s2}$. For the branch with $\gamma=1$, $r_{s1}$ and $r_{s2}$ represent the inner horizon and the event horizon, respectively. For the branch with $\gamma=-1$ and even $n$, $r_{s1}$ and $r_{s2}$ represent the event horizon and the cosmological horizon, respectively.

\subsection{A brief discussion about the topology}

Given that in our solutions the mass function is such that the spacetime behaves differently both near and far from the origin, it is of physical interest to test the topology of the spacetime at different locations. In this regard, initially, in reference \cite{Borde:1996df}, using the analogy with the Reissner-Nordström spacetime, conditions for the so-called Borde's theorem were established, which leads to the fact that regular black holes with spherical symmetry (and consequently compact on the event horizon) in their cross-section are associated with a change in topology. Subsequently, in the references \cite{Bargueno:2020ais,Melgarejo:2020mso,Bargueno:2021fus} the topology of four-dimensional regular black holes with (A)dS cores and other types of cores and with spherical symmetry in their cross-section was studied. In reference \cite{Bargueno:2020ais}, it was found that the geometry of spacelike slices of spherically symmetric and (locally) static regular solutions is $S^3$ for a de Sitter core and $H^3$ for an AdS core. In that reference, it was pointed out that the regular black holes with dS core satisfying Borde's theorem \cite{Borde:1996df} has the topology of $S^3$ in their cores, assuming the compact slice is smooth and simply connected. On the other hand, there is a transition between the topology of the core and the topology far from it, from $S^3$ to $\mathcal{R} \times S^2$. Related to the latter, in reference \cite{Melgarejo:2020mso}, a way of representing the spacetime of a regular black hole was addressed, providing an identification of the regions corresponding to the origin in the usual Penrose diagram of Reissner-Nordström. This allows the spacetime to be described as causally simple towards the future and geodesically complete towards the future. It is also shown that all the requirements of Borde's theorem are satisfied by imposing the null energy condition. Therefore, there must be some compact slice in the causal future of the eventually trapped surface. This spacelike identification implies that the topology of the slices in the inner region of the regular black hole is $S^3$, where the metric $ds^2 = r_- d\chi^2 + r^2(\chi) d\Omega^2$ is such that $r(\chi)$ fluctuates between $0$ and $r_-$. This type of identification is also consistent with the aforementioned transition in topology.

In this subsection, we will have a brief discussion on the topology associated with the cases described in the previous subsections.

\subsubsection{Branch with $\gamma = 1$ for both even and odd $n$ for $d>2n+1$}

As mentioned before, the case where $d = 2n + 1$ does not represent a black hole; therefore, the case of interest corresponds to $d > 2n + 1$. As we can observe in equation \eqref{NucleoDeSiter}, the core corresponds to a dS space. In order to test the topology, we will follow a strategy analogous to that in reference \cite{Bargueno:2020ais}. Since we are interested in testing whether the topology of the core corresponds to a sphere $S^{d-1}$, we will assume that the cross section of a coordinate system in any spacelike slice is constant and positively normalized to unity, $\gamma = 1$. However, a more detailed study, such as the one in reference \cite{Bargueno:2020ais}, may be required, which goes beyond the scope of this work. Such a coordinate system can be written as

\begin{equation} \label{Lamina}
    ds^2= \frac{dr^2}{\lambda(r)}+ r^2 d\Omega_{d-2}. 
\end{equation}
where $\lambda(r)$ is given by \eqref{NucleoDeSiter}. This last equation can be rewritten as:
\begin{equation}
    \lambda(r)= 1-\frac{r^2}{l_{eff}}=1-\frac{R_{dS}}{d(d-1)}r^2=1- \frac{2}{(d-2)}\frac{\Lambda_{eff}}{(d-1)}r^2
\end{equation}
where $l_{eff}$, $R_{dS}$, and $\Lambda_{eff}$ represent the effective radius, Ricci scalar, and effective cosmological constant for the de Sitter spacetime, respectively. Intuitively motivated by reference \cite{Bargueno:2020ais}, the following coordinate transformation is proposed:
\begin{equation}
    1-\frac{R_{dS}}{d(d-1)}r^2 = \cos^2(\chi) \Rightarrow r= \left ( \frac{d(d-1)}{R_{dS}} \right)^{1/2} \sin (\chi)
\end{equation}
with this change of coordinate, the line element \eqref{Lamina} is:
\begin{equation}
    ds^2= \frac{d(d-1)}{R_{dS}} \left ( d\chi^2+ \sin^2(\chi) d\Omega_{d-2} \right)
\end{equation}
And therefore, the topology of the spacelike slices is described by either $S^{d-1}$ (corresponding to a de Sitter core with $R_{dS} > 0$). On the other hand, we can check in equation \eqref{ConvergenciadS} and figure \ref{convergency} that, given the form of our mass function, the space time far from the origin behaves as the vacuum solution whose topology corresponds to $\mathcal{R} \times S^{d-2}$. Thus, we can infer that in our case, there is a transition in topology from an $S^{d-1}$ near the core to a topology of $\mathcal{R} \times S^{d-2}$ far from it. This result in $4D$ reduces to that in reference \cite{Bargueno:2020ais}

\subsubsection{Branch with $\gamma = -1$ for even $n$}

First of all, as pointed out previously, the case $d = 2n + 1$ does not have an inner horizon associated with a potentially unstable core.
On the other hand, for $d > 2n + 1$, there is a black hole with both an event horizon and a cosmological horizon. As we can observe in equation \eqref{NucleoAdS}, the solution has an AdS core. This feature makes it intriguing to test whether or not there is a topological transition analogous to the previous case, where now the cross-section corresponds to a hyperboloid spacetime. However, the fact that in our case the cross-section corresponds to a compact hyperboloid, together with the existence of a cosmological horizon, implies that the analogous identification with the Reissner-Nordström spacetime is not direct and requires a deep study, which is beyond the scope of this work and could be addressed in future studies.

\section{Temperature}

As mentioned above, the nonvanishing temperature black hole solutions can be computed at $r=r_+$. In this case, the thermodynamics can be analyzed considering the space between $r_+ < r < \infty$.  Evaluating the equations of motion \eqref{ttcomponente} at the horizon yields the following expression for the temperature

\begin{equation} \label{TemperaturaGamma}
T=  \frac{1}{4 \pi n \gamma^{n-1}} \left (  \frac{d-2n-1}{r_+} - \frac{2}{d-2} \rho (r_+) r_+^{2n-1} \right )
\end{equation}
where the mass parameter present in the energy density, equations \eqref{DensidadDeEnergia} and \eqref{ADensidadDeEnergia}, corresponds to that in equation \eqref{rplusSol}.

Since the mass parameter $M$ can be expressed as a function of $r_+$, $r_{++}$ or $r_-$ (as shown in Equation \eqref{rplusSol} and illustrated in Figures \ref{M1d=2n+1}, \ref{MImpar}, and \ref{Mpar}), this implies the existence of a minimal mass parameter, denoted $M(r_*)$, where $r_+ = r_- = r_*$ (or where $r_+ = r_{++} = r_*$) , with $r_*$ being the radius at which the system's temperature vanishes. This can also be observed in the parameter space using the triple product chain rule:

\begin{equation} \label{TripleChainRule}
    \delta f =0= \frac{\partial f}{\partial r_+} dr_+ + \frac{\partial f}{\partial M} dM \rightarrow \frac{\partial f}{\partial r_+} \frac{\partial r_+}{\partial M} \frac{\partial M}{\partial f} =-1 \rightarrow T \sim \frac{\partial f}{\partial r_+} \sim -\frac{\partial f}{\partial M} \frac{\partial M}{\partial r_+}
\end{equation}

Thus, since there is a minimum at $\frac{\partial M}{\partial r_+}$, where the plots show that the inner and outer horizons coincide, at that point $T = 0$. We can also note that: For the branch with \( \gamma = 1 \), \( -\frac{\partial f}{\partial M} > 0 \) and \( \frac{\partial M}{\partial r_+} \ge 0 \), so the sign of the temperature is always positive. For the branch with \( \gamma = -1 \), \( -\frac{\partial f}{\partial M} < 0 \) and \( \frac{\partial M}{\partial r_+} \le 0 \), so the sign of the temperature is also always positive.

To find an expression for the minimum value of $M$ where $T=0$, we can notice that the temperature can also be expressed as an equation for $\frac{r_s}{a}$
\begin{equation}\label{T+}
 T_+ =  \frac{1}{4\pi} \left. \frac{d\mu}{dr} \right|_{r=r_+} = {\frac{1}{\gamma}} \left (-\frac{1}{2 \pi  r_{+}} +\frac{{}_1F_{1} \left(\left[n+1\right],\left[n+2\right],-\left(\frac{r_+}{a}\right)^{\frac{d-1}{n}}\right) \left(\frac{r_+}{a}\right)^{\frac{d-1}{n}+2 n} \left(d-1\right)}{2 \pi   r_{+} n^{2} \left(n+1\right)} \right)
\end{equation}
where the value of \( r_+ \) corresponds to the largest solution of \( \mu(r) = 0 \) from Equation \eqref{SolucionGamma1} for the branch \( \gamma = 1 \), and to the smallest solution of \( \mu(r) = 0 \) from Equation \eqref{SolucionGamma-1} for the branch \( \gamma = -1 \).

Eq.(\ref{T+}) shows that the temperature $T_+$ can vanish, \textit{i.e.}, there are values of $M/a^{d-2n-1}$ for which $T_+=0=T_-$ (or $T_+=0=T_{++}$). Because the relation between $M/a^{d-2n-1}$ and $r_+$ is not analytic, it is much simpler to determine $T_+=0$, or equivalently $r_+=r_-=r_*$ (or $r_+=r_{++}=r_*$ ), using the condition
\begin{equation}
\frac{{}_1F_{1} \left(\left[n+1\right],\left[n+2\right],-\left(\frac{r_*}{a}\right)^{\frac{d-1}{n}}\right) \left(\frac{r_*}{a}\right)^{\frac{d-1}{n}+2 n} \left(d-1\right)}{n^{2} \left(n+1\right)} = 1.
\end{equation}
One can notice that this is an equation for $\frac{r_*}{a}$ and thus given $d$ and $n$ its solution can be computed at least numerically. This implies that the minimum value of the (normalized) mass, says $M_*$, is given by
\begin{equation}\label{minimalMass}
 \frac{M_{\text{min}}}{a^{d-2n-1}} = M_* = \frac{d-1}{2n(n+1)} \left(\frac{r_*}{a} \right)^{\frac{d-1}{n}}
\end{equation}

\subsection{branch with \(\gamma = 1\)}

As previously mentioned, the case $d = 2n + 1$ does not represent a black hole, as it lacks an event horizon. Therefore, we will examine the case where $d > 2n + 1$ below.

\subsubsection{\bf $d>2n+1$} 

\begin{figure}[h]
    \centering
    \includegraphics[scale=.4]{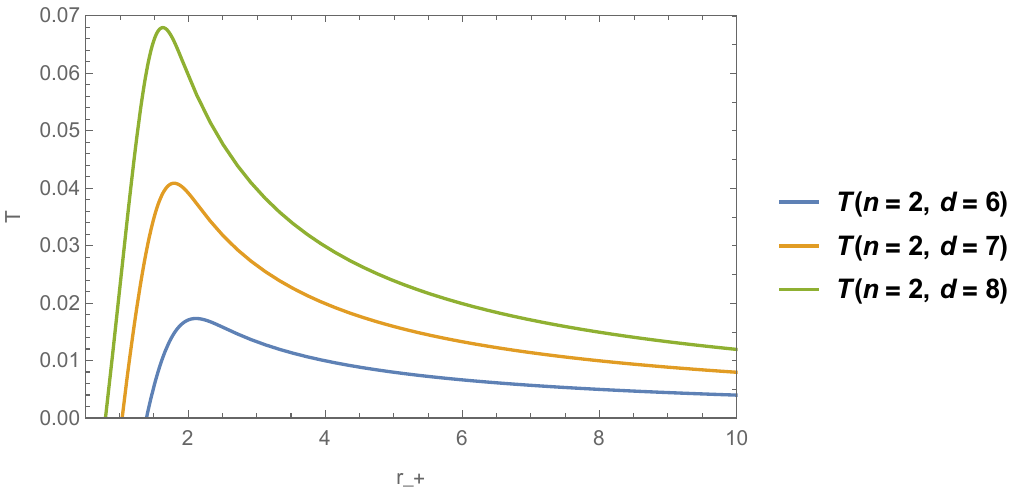}
    \includegraphics[scale=.4]{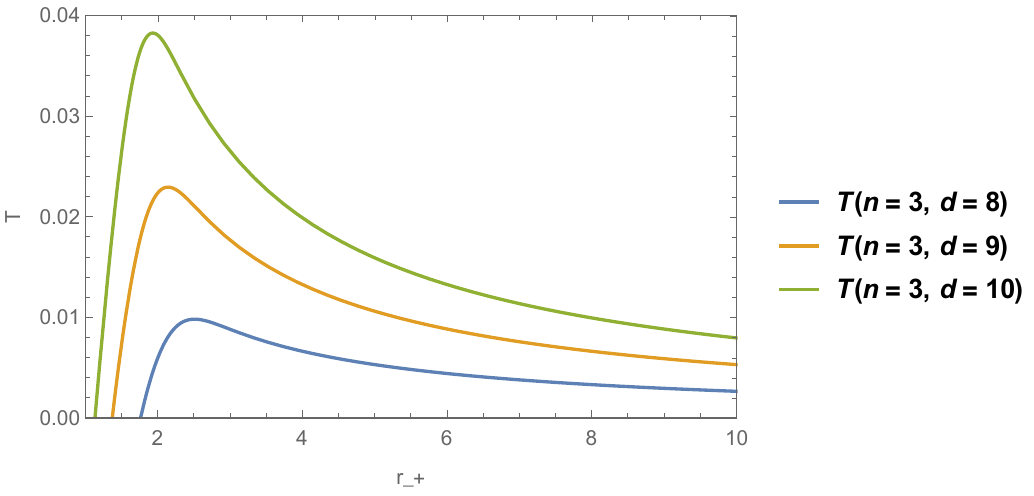}
    \caption{Left side: Temperature for $n=2$, $d=6$, $a=1$. Right side: Temperature for $n=3$, $d=8$, $a=1$.}
    \label{FigTemperatura1}
\end{figure}

We can observe the behavior in Figure \ref{FigTemperatura1}. It is straightforward to verify that this behavior is generic for other values of $n$ and $d$.

We can verify that the point \( r_* \), where the mass parameter reaches a minimum (i.e., the internal and black hole horizons coincide), is the same point where the temperature vanishes. Hence, the temperature reaches to zero at the extremal black hole. As we will discuss later, in this case, the zero-temperature point is associated with a black remnant, which refers to what remains after evaporation has ceased. 

From right to left in the figure \ref{FigTemperatura1}, we observe that: first, being the derivative \( dT/dr_+ < 0 \), the temperature increases as the horizon decreases, reaching a maximum at \( r = r_{max} \). At this maximum point, \( dT/dr_+ = 0 \). Then, with \( dT/dr_+ > 0 \), the temperature decreases along with the horizon until \( T = 0 \) at \( r_* \). Below, we will discuss the physical implications of this temperature behavior in the context of radial evolution and evaporation.

\subsection{branch with \(\gamma = -1\)}

\subsubsection{\bf $d=2n+1$} 
We can observe the behavior on the left side of Figure \ref{FigTemperatura-1}. It is straightforward to verify that this behavior is generic for other values of $n$ and $d$.
As mentioned above, for $d=2n+1$ with $\gamma=-1$ and $n$ even, the geometry corresponds to that of a regular black hole. Although the figure shows only a fraction of the domain, we can verify that the derivative \( \frac{dT}{dr_+} \) is always negative. Thus, we speculate that as the horizon radius increases, the derivative asymptotically approaches zero, with the temperature also approaching zero. This suggests that as the event horizon expands, a finite residual value might be reached for the latter. Below, we will discuss the physical consequences for the radial evolution and evaporation of this behavior.

\subsubsection{\bf $d>2n+1$} 

\begin{figure}[h]
    \centering
    \includegraphics[scale=.4]{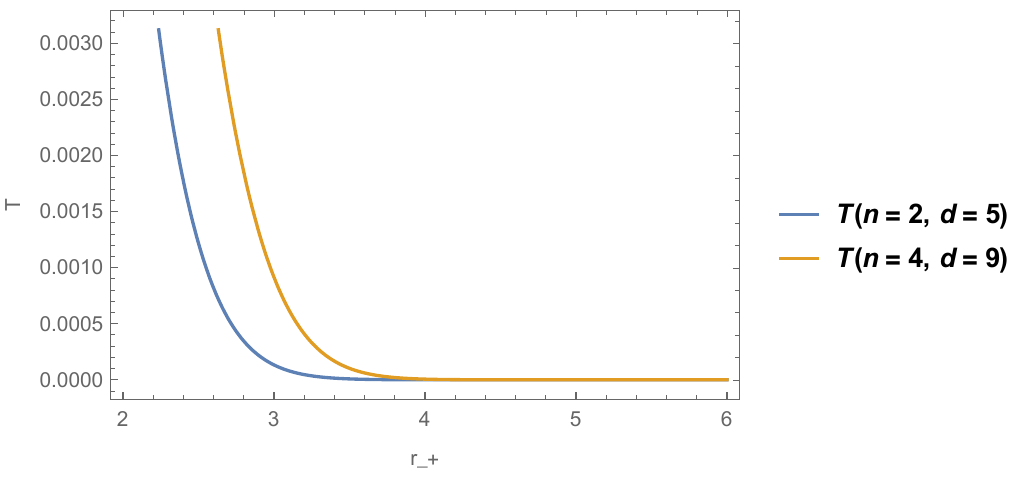}
    \includegraphics[scale=.4]{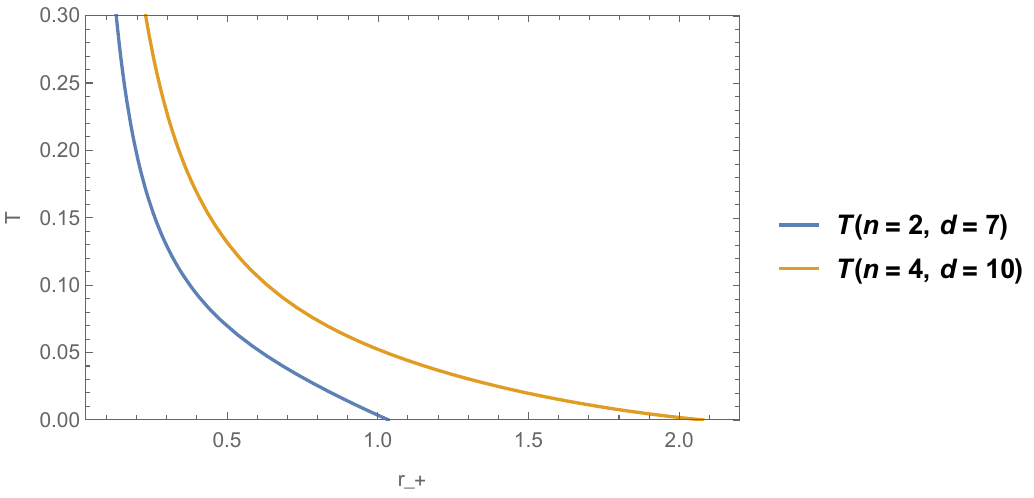}
    \caption{Left side: Temperature for $n=2$, $d=5$, $a=1$. Right side: Temperature for $n=4$, $d=9$, $a=1$.}
    \label{FigTemperatura-1}
\end{figure}

We can observe the behavior on the right side of Figure \ref{FigTemperatura-1}. Similarly to the previous case, although the figure shows only a fraction of the domain, we can verify that the derivative \( \frac{dT}{dr_+} \) is always negative. Thus, if the horizon radius decreases, the temperature always increases, suggesting that the temperature would need to increase to infinity for the horizon radius to completely vanish. On the other hand, we observe that as the horizon radius increases, the temperature decreases until it reaches a null value, \( T = 0 \). This also occurs at the minimum value of the mass parameter; however, unlike the \(\gamma = 1\) branch, this happens at the point where the event and cosmological horizons coincide. In other words, the event horizon could expand until it intersects with the cosmological horizon. Below, we will discuss the physical implications of this temperature behavior in the context of radial evolution and evaporation.

\section{Entropy and the first law}

Given the form of the Lagrangian and the family of solutions considered it is straightforward to define an entropy at the event horizon. For this one can use the methodology showed in reference \cite{Wald:1993nt}
\begin{equation}
    S_+ = \int_{\mathcal{H}} \frac{\partial L}{\partial R^{01}_{\hspace{2ex}01}} \approx (\gamma)^{n-2} r^{d-2n} \Sigma_{\gamma}
\end{equation} 
which coincides with the known expression for the vacuum non-regular black holes \cite{Cai:2006pq}. One can notice that this is an always increasing function of $r_+$. Two important caveats exist. First, while the expression is the same, even though the values are extremely close, the values of $r_+$ differ between the solution above and the solution in \cite{Cai:2006pq}. The second, as the solution exists for $\gamma=1$ for any $n$ but for $\gamma=-1$ only if $n$ is even, the expression of the entropy is the same in both cases. 

Following Wald's construction it is straightforward to confirm that, under an evolution of the parameters of the solution,  
\begin{equation}\label{firstlaw}
    T_+ \delta S_+ = \delta M
\end{equation}
is satisfied. Even though this is the standard result, it must be considered that in this case, $T_+$ can vanish. Moreover, it seems like the parameter $a$ is completely absent, but this is only apparent as the value of $r_+$ depends on $a$. 

It must be emphasized in this point that Eq.(\ref{firstlaw}) does not provide the whole scenario to understand the evolution of the system.

\section{Radial evolution and Heat capacity}

It is important to emphasize that a discussion of thermodynamics is only possible in the presence of an event horizon. In this case, the mass \( M \) can be expressed as a function of \( r_s=r_+ \), as given by Equation \eqref{MasaHipergeometrica}.

Afterwards, the heat capacity of these solutions can be computed as
\begin{equation}\label{HeatCapacity}
    C(r_+) = \frac{\partial M}{\partial T} = \frac{dM}{dr_+} \left(\frac{dT}{dr_+}\right)^{-1}
\end{equation}

First, we will outline an analytical expression and then analyze the radial evolution and evaporation for each of the cases of interest. From Eq.(\ref{T+}),
\begin{eqnarray}
  \frac{d}{dr_+} T(r_+)  &=& \frac{1}{\gamma} \bigg (\frac{1}{2 \pi  a \left(\frac{r_+}{a}\right)^{2}}-\frac{{}_1F_{1} \left(\left[n+2\right],\left[n+3\right],-\left(\frac{r_+}{a}\right)^{\frac{d-1}{n}}\right) \left(\frac{r_+}{a}\right)^{\frac{d-1}{n}} \left(d-1\right)^{2} \left(\frac{r_+}{a}\right)^{\frac{d-1}{n}+2 n}}{2 \left(n+2\right) n^{3} \left(\frac{r_+}{a}\right)^{2} \pi  a} \nonumber\\
   &+& \frac{{}_1F_{1}\left(\left[n+1\right],\left[n+2\right],-\left(\frac{r_+}{a}\right)^{\frac{d-1}{n}}\right) \left(\frac{r_+}{a}\right)^{\frac{d-1}{n}+2 n} \left(\frac{d-1}{n}+2 n\right) \left(d-1\right)}{2 \left(\frac{r_+}{a}\right)^{2} \pi  a \,n^{2} \left(n+1\right)} \label{dT}\\
   &-& \frac{{}_1F_{1}\left(\left[n+1\right],\left[n+2\right],-\left(\frac{r_+}{a}\right)^{\frac{d-1}{n}}\right) \left(\frac{r_+}{a}\right)^{\frac{d-1}{n}+2 n} \left(d-1\right)}{2 \pi  a \left(\frac{r_+}{a}\right)^{2} n^{2} \left(n+1\right)}\nonumber \bigg )
\end{eqnarray}
and
\begin{eqnarray}
   \frac{dM}{dr_+}&=& \frac{a^{d-2 n-1} n {}_1F_{1} \left(\left[n+1\right],\left[n+2\right],-\left(\frac{r_+}{a}\right)^{\frac{d-1}{n}}\right) \left(\frac{r_+}{a}\right)^{\frac{d-1}{n}} \left(d-1\right)}{2 {}_1F_{1}\left(\left[n\right],\left[n+1\right],-\left(\frac{r_+}{a}\right)^{\frac{d-1}{n}}\right)^{2} \left(\frac{r_+}{a}\right)^{d-1} \left(n+1\right) \left(\frac{r_+}{a}\right)} \nonumber\\
 &-& \frac{a^{d-2 n-1} n \left(d-1\right)}{2 {}_1F_{1} \left(\left[n\right],\left[n+1\right],-\left(\frac{r_+}{a}\right)^{\frac{d-1}{n}}\right) \left(\frac{r_+}{a}\right)^{d-1} \left(\frac{r_+}{a}\right)} \label{dM}
\end{eqnarray}
where again the value of \( r_+ \) corresponds to the largest solution of \( \mu(r) = 0 \) from Equation \eqref{SolucionGamma1} for the branch \( \gamma = 1 \), and to the smallest solution of \( \mu(r) = 0 \) from Equation \eqref{SolucionGamma-1} for the branch \( \gamma = -1 \).

There are two important features to notice. First $dT/dr_+$ can vanish pointing out potential phase transitions. Second, the values of $r_+$ where this occurs are not obviously connected with $r_*$, the value that defines the minimal mass $M_*$ and $T_+=0$. Moreover, $dM/dr_{+}$ can vanish as well, in another seemly independent value of $r_+/a$, pointing out regions where the heat capacity goes smoothly from $C>0$ to $C<0$. This usually implies a rich \textit{adiabatic} evolution of the solutions.

Of lesser relevance but still worthwhile to mention is that, even though equations (\ref{dT}) and (\ref{dM}) may look quite cumbersome, in fact once expressed for a given $d$ and $n$ they simplify greatly.

For our analysis, we will consider the following: The heat capacity will be utilized to study the thermodynamic evolution of the black hole. In this work, a positive heat capacity indicates that when the temperature decreases (increases), the black hole emits (absorbs) thermal energy, thus \( dM < 0 \) (\( dM > 0 \)) in the black hole, in order to reach thermodynamic equilibrium with the external environment, i.e., the black hole is stable. Conversely, a negative heat capacity represents that, if the temperature increases (decreases), the black hole also emits (absorbs) thermal energy toward the external environment, i.e., the black hole is unstable. A second-order phase transition is characterized by a change in the sign of the heat capacity. 

\subsection{branch with \(\gamma = 1\)}

For $d>2n+1$, as mentioned earlier, since \( r_+ \) represents the outermost horizon, \( r_+ > r_- \). In Figures \ref{MImpar} and \ref{Mpar}, we can observe that the derivative \( \frac{dM}{dr_+} \ge 0 \). Thus, the sign of the specific heat depends solely on the sign of the derivative \( \frac{dT}{dr_+} \). 

Following Figure \ref{FigTemperatura1}, we propose the following interpretation from right to left: Starting from the right of \( r_{\text{max}} \), the specific heat is negative, so energy is emitted into the environment while the temperature increases and the horizon radius contracts. Subsequently, at the point \( r_{\text{max}} \), the specific heat diverges and a phase transition occurs from an unstable black hole to a stable black hole. Later, to the left of \( r_{\text{max}} \), the specific heat is positive. Thus, while the horizon radius continues to contract, the black hole continues to emit energy and the temperature decreases. Finally, at the point where the temperature goes to zero, the specific heat also goes to zero, and the event horizon halts its contraction. Thus, once the contraction stops at the zero-temperature point, a black remnant could form, referred to as what is left behind once evaporation ceases \cite{Adler:2001vs}. Some references have explored the possibility that evaporation would stop once the horizon radius contracts to a value close to the Planck length, a phenomenon that might be linked to the emergence of quantum effects at this scale \cite{Estrada:2023cyx}.

\subsection{branch with \(\gamma = -1\)}

\subsubsection{\bf $d=2n+1$} 

\begin{figure}[h]
    \centering
    \includegraphics[scale=.4]{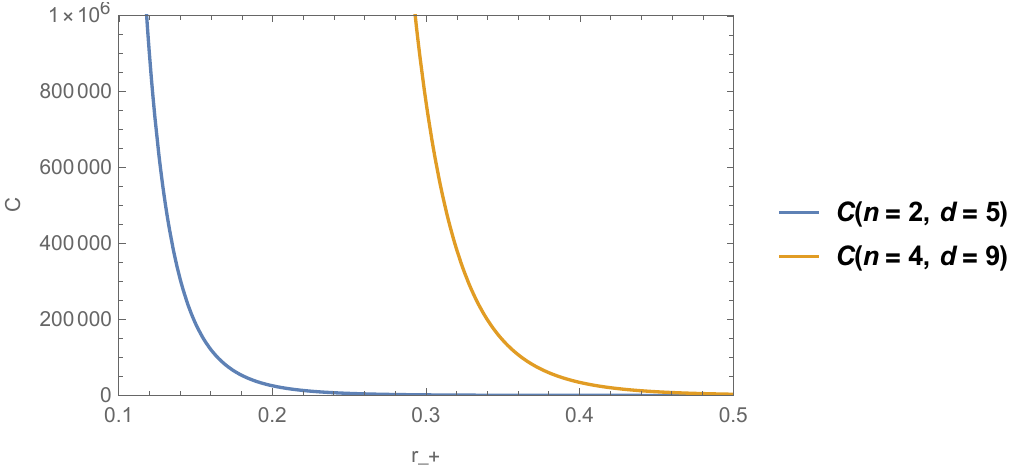}
       \caption{Left side: Specific heat for $n=2$, $d=6$, $a=1$. Right side: Specific heat for $n=4$, $d=9$, $a=1$}
    \label{C-12n+1}
\end{figure}

In this case, as mentioned earlier, since \( r_+ \) represents the only horizon corresponding to the event horizon, we can check the left side of Figure \ref{M1d=2n+1}, for \( n = 2,4 \), that the derivative \( \frac{dM}{dr_+} < 0 \). On the other hand, we can see on the left side of Figure \ref{FigTemperatura-1} that the sign of the derivative \( \frac{dT}{dr_+} \leq 0 \). Thus, the specific heat is always positive. Since interpreting the behavior solely from the graphs of the parameter \( M \) and temperature does not seem straightforward, we outline the behavior of the specific heat, given by Equation \eqref{HeatCapacity}, in Figure \ref{C-12n+1}.

Moving from left to right in figure \ref{C-12n+1}, we observe that there is a value of the event horizon where an asymptote of zero is reached. In other words, this point, where the specific heat approaches zero, can be associated with a remnant. Thus, speculating, as the black hole horizon expands and cools, it transfers energy to the environment, halting evaporation once the mentioned asymptote is reached.

\subsubsection{\bf $d>2n+1$} 

In the literature, it has been theoretically explored that the cosmological horizon could also have a temperature \cite{Aros:2008ef,Kubiznak:2015bya,Estrada:2019qsu}. This is beyond the scope of our work. However, this fact might provide some insights into the radial evolution of the system. From Equation \eqref{TripleChainRule}:
\begin{equation}
 dM= - \left ( \frac{1}{4\pi} \frac{\partial f}{\partial r_+} \right)  dr_+ = - \left ( \frac{1}{4\pi} \frac{\partial f}{\partial r_{++}} \right)  dr_{++} 
\end{equation}

First, we can check that the quantity \( - \frac{\partial f}{\partial M} < 0 \) and \( \frac{\partial M}{\partial r_{++}} > 0 \) at the cosmological horizon (see Figure \ref{Mpar}). Thus, following the triple chain rule, analogous to Equation \eqref{TripleChainRule}, it is customary to define the temperature at the cosmological horizon as \( T_{++} \sim - \frac{\partial f}{\partial r_{++}} \sim \frac{\partial f}{\partial M} \frac{\partial M}{\partial r_{++}} > 0 \). Thus, we can rewrite the last equation as:

\begin{equation} \label{EvolucionRadialDeSitter}
 dM= - T_+  dr_+ = T_{++}  dr_{++} 
\end{equation}

First, we can observe in Figure \ref{Mpar} that $\frac{\partial M}{\partial r_+} < 0$, noting that in this case, the event horizon corresponds to the smallest of the horizons, $r_+ < r_{++}$. On the other hand, in the right side of Figure \ref{FigTemperatura-1}, we can see that $\frac{\partial T}{\partial r_+} < 0$. Therefore, the specific heat at the event horizon is always positive.

Assuming that the event horizon expands from left to right while the temperature decreases until it reaches $T = 0$ on the right side of Figure \ref{FigTemperatura-1}, our interpretation is as follows: From the latter equation \eqref{EvolucionRadialDeSitter}, we can deduce that since the specific heat at the event horizon is positive, if the object emits energy to its surroundings ($dM < 0$), the event horizon expands ($dr_+ > 0$), and the cosmological horizon contracts ($dr_{++} < 0$). Once both horizons reach the point $M(r_*)$ where the temperature becomes zero, the evaporation process slows down. In other words, the physical system reaches equilibrium when both horizons coincide. It is worth mentioning that this process is analogous to the evaporation of a black hole with a positive cosmological constant and a spherically symmetric cross section. However, in our case, this process occurs without the presence of a cosmological constant, and the transversal section represents a hyperboloid.

\section{Discussion and summarize}

In this work, we have constructed a new family of regular black hole solutions for pure Lovelock gravity. The energy density was developed using arguments analogous to those of the Dymnikova model: the energy density encodes the gravitational information of the vacuum case through the Kretschmann scalar. Near the radial origin, where tidal forces and the gravitational tension in the vacuum case diverge, the tidal forces in our model become finite due to the specific form of the energy density. This can be verified as both the geometry and the curvature invariants in our model remain finite. Unlike the Dymnikova model, our energy density varies with the power \(n\) of the Riemann tensor in the Lovelock theory and with the number of dimensions. Additionally, we have considered cases where the cross-section in the vacuum case corresponds to a hyperboloid.

Speculatively, our model might capture quantum effects through gravitational tension, which is proportional to the square root of the Kretschmann scalar in the vacuum case. In this context, some references, which we mention further below, suggest that, under these assumptions, vacuum polarization could occur within the resulting gravitational field. Consequently, these reference proposes an analogy between the ratio of pair production in the Schwinger effect and the energy density of the Dymnikova model, denoted as $\Gamma \sim \rho_{\text{Dymnikova}}$. This idea has been proposed in works \cite{DymnikovaS1996,Ansoldi:2008jw}, and more recently in references \cite{Estrada:2023pny,Alencar:2023wyf} for general relativity in \(4D\). In our model, the vacuum corresponds to pure Lovelock gravity, characterized by either spherical or hyperbolic transverse geometry. However, a more in-depth study is required, which is beyond the scope of this work.

We have found two cases of interest: the branch with $\gamma=1$, which has a spherically symmetric cross-section for both even and odd $n$; and the branch with $\gamma=-1$, which has a hyperbolic cross-section for even $n$. 

{\bf For the branch $\gamma=1$}, whose transversal section corresponds to a sphere, we can highlight the following:

\begin{itemize}
    \item For $d=2n+1$, if the mass parameter satisfies $M>M_*$, the solution can be interpreted as a static cosmology, since in this case there is a cosmological horizon present.
    \item For $d>2n+1$, when $M=M^*$, the geometry represents a regular extremal black hole where the inner and event horizons coincide and where the temperature is zero. 
    \item For $d>2n+1$, when $M>M^*$, the geometry corresponds to a regular black hole with both an external and an internal horizon. This solution has a dS core. One very interesting feature of our solution is how quickly it converges to the non-regular vacuum solution \cite{Cai:2006pq}. In fact, it seems that the two functions match within the region $r_{-}<r<r_{+}$. In reality, the separation between the two functions becomes very small rapidly, making it nearly impossible to distinguish between the solutions outside the outermost horizon.
    \item We have also analyzed the radial evolution of the temperature, which provides insights into the evaporation process. The temperature reaches a maximum at \( r_{\text{max}} \). Starting from the right of \( r_{\text{max}} \), the specific heat is negative, so energy is emitted into the environment while the temperature increases and the horizon radius contracts. At \( r_{\text{max}} \), the specific heat diverges, and a phase transition occurs from an unstable black hole to a stable one. Moving to the left of \( r_{\text{max}} \), the specific heat becomes positive. Thus, while the horizon radius continues to contract, the black hole continues to emit energy, causing the temperature to decrease. Finally, at the point where the temperature approaches zero, the specific heat also drops to zero, and the event horizon halts its contraction. This occurs in the extremal case. Thus, when the contraction stops at the zero-temperature point, a black hole remnant is formed, which is what remains once evaporation ceases \cite{Adler:2001vs}. Some references have explored the possibility that evaporation would stop once the horizon radius contracts to a value close to the Planck length, a phenomenon that might be linked to the emergence of quantum effects at this scale \cite{Estrada:2023cyx}.
\end{itemize}

{\bf For the branch $\gamma=-1$ with even $n$ and with $d=2n+1$}, whose transverse section corresponds to a hyperboloid, we can highlight the following:

\begin{itemize}
    \item For $M>M_*$, the solution represents a regular black hole with an event horizon and no inner horizon. This feature is particularly interesting because, in recent years, the literature has discussed both the instability of an inner horizon with a dS core due to mass inflation \cite{Carballo2023}, as well as the possibility that Hawking radiation might mitigate the instability associated with the inner horizon and dS core \cite{Bonanno:2022jjp}.
    \item As we move radially outward from $r_+ \to 0$, the derivative \( \frac{dT}{dr_+} \) is negative, approaching \( T=0 \) asymptotically. The specific heat is positive and also approaches zero asymptotically. The point where the temperature and specific heat approach zero can be associated with a remnant. Thus, speculating, as the black hole expands and cools, it transfers energy to the environment, halting evaporation once the aforementioned asymptote is reached.
\end{itemize}

{\bf For the branch $\gamma=-1$ with even $n$ and with $d>2n+1$}, whose transverse section corresponds to a hyperboloid, we can highlight the following:

\begin{itemize}
    \item For \( M < M(r_*) \), the solution has no horizons and seems to lack physical interest .
    \item For \( M > M(r_*) \), there is an event horizon and a cosmological horizon. The presence of the cosmological horizon prevents the computation of energy using Noether's charge, meaning the parameter \( M \) is not directly associated with the mass. However, through thermodynamic arguments, a local definition of mass variation can be established, which depends on \( M \) \cite{Aros:2008ef}. A remarkable feature of this solution is the presence of both an event horizon and a cosmological horizon without the need for a positive cosmological constant, as seen in the Schwarzschild-de Sitter case.
    \item For \( M = M(r_*) \), this represents the extremal case where the event horizon and the cosmological horizon coincide. This geometry is analogous to the Nariai solution \cite{Nariai1999}, which arises from the evolution of the Schwarzschild-de Sitter black hole. At this point, the temperature becomes zero.
    \item As we move radially outward from $r_+ \to 0$, the derivative \( \frac{dT}{dr_+} \) is negative, reaching \( T=0 \) in the extremal case where $r_+ = r_{++}$. From both analytical and graphical analysis, we can deduce that since the specific heat at the event horizon is positive, if the object emits energy to its surroundings ($dM < 0$), the event horizon expands ($dr_+ > 0$), and the cosmological horizon contracts ($dr_{++} < 0$). Once both horizons reach the point $M(r_*)$, where the temperature becomes zero, the evaporation process slows down. In other words, the physical system reaches equilibrium when the horizons coincide. It is worth noting that this process is analogous to the evaporation of a black hole with a positive cosmological constant and a spherically symmetric cross-section. However, in our case, this process occurs without the presence of a cosmological constant, and the transverse section corresponds to a hyperboloid.
\end{itemize}

Moreover, it is worth mentioning that, following references \cite{Borde:1996df,Bargueno:2020ais,Melgarejo:2020mso,Bargueno:2021fus}, we have established that for the branch with a spherically symmetric cross-section for $d > 2n + 1$, near the de Sitter core, the topology of the spacelike slices is described by an $S^{d-1}$ sphere (corresponding to a de Sitter core with $R_{dS} > 0$). On the other hand, given the form of the mass function, the spacetime far from the origin behaves as the vacuum solution, whose topology corresponds to $\mathcal{R} \times S^{d-2}$. Thus, we can infer that in this case, there is a transition in topology from an $S^{d-1}$ near the core to a topology of $\mathcal{R} \times S^{d-2}$ far from it. This result in $4D$ reduces to that in reference \cite{Bargueno:2020ais}.

\acknowledgments

This work of RA was partially funded through FONDECYT-Chile 1220335. Milko Estrada is funded by the FONDECYT Iniciaci\'on Grant 11230247.

\bibliography{mybib.bib}

\end{document}